\documentclass[10pt,a4paper,twocolumn,conference]{IEEEtran}
\usepackage[margin=1.9 cm]{geometry}
\usepackage{cite}	% citation
\usepackage{graphicx} % grafics, floats (figs)
\usepackage[latin1]{inputenc} % language
\usepackage[T1]{fontenc} % language and fonts
\usepackage{amsmath,amsfonts,amsbsy,amssymb} % mathematical lybraries
\usepackage{mathabx} 
\usepackage{amssymb} 
\usepackage{amsmath}
\usepackage{amsthm}	
\usepackage{mathrsfs}
\usepackage[nolist]{acronym} % acronym lybrary
\usepackage{tabularx} % tables
\usepackage{multirow}
\usepackage{wasysym}
\usepackage{float}
\usepackage{url}
\usepackage{color} % colors (fonts and edting)
\usepackage{graphicx}
\usepackage[caption=false]{subfig}
\usepackage{lipsum} 
\usepackage[keeplastbox]{flushend}
\usepackage{enumitem} % special package for enumeration
\usepackage[font=small,skip=0pt]{caption}
\usepackage{caption}
\DeclareCaptionLabelFormat{mylabel}{#1#2.\hspace{1ex}}
\begin{document}
	
	% - - - - - - - - - - - - - - - - - - - - - - - - - - - - - - - - - - - - - - - - - -
	\title{In-Band Pilot Overhead in Ultra-Reliable Low Latency Decode and Forward Relaying}
	\author{Parisa Nouri\IEEEauthorrefmark{0}, Hirley Alves\IEEEauthorrefmark{0}, Richard Demo Souza\IEEEauthorrefmark{1}, and Matti Latva-aho \IEEEauthorrefmark{0}\\
		\IEEEauthorblockA{
			\IEEEauthorrefmark{0}Centre for Wireless Communications (CWC), University of Oulu, Finland\\
		}
		\IEEEauthorblockA{
			\IEEEauthorrefmark{1} Federal University of Santa Catarina (UFSC), Brazil\\
		}
		%	\IEEEauthorblockA{
		%		\IEEEauthorrefmark{3}Centre for Wireless Communications (CWC), University of Oulu, Finland\\
		%	}
		
		\{Parisa.Nouri, Hirley.Alves, and Matti.Latva-aho\}@oulu.fi,
		\IEEEauthorrefmark{1}richard.demo@ufsc.br
	}
	%
	% - - - - - - - - - - - - - - - - - - - - - - - - - - - - - - - - - - - - - - - - - -
	\maketitle
	
	% - - - - - - - - - - - - - - - - - - - - - - - - - - - - - - - - - - - - - - - - - -
	\begin{abstract}
		In URLLC the performance of short message communications highly depends on the training sequence length due to the stringent latency and reliability requirements. In this paper, we study the performance of cooperative and non-cooperative transmissions under imperfect channel estimation and Rayleigh fading for URLLC. We assume a peak power constraint on pilot symbols in addition to the average power constraint which is used for comparison purposes.  We obtain the optimal training length as a function of blocklength and power constraint factor to meet the URLLC requirements. Moreover, the simulation results show the impact of pilot overhead on reliability, latency, and goodput of cooperative communications compared to point-to-point transmission.
	\end{abstract}
	
\vspace{0.75mm}
\section{Introduction}

The fifth generation (5G) of the cellular systems inherently handles three essential use cases, namely enhanced mobile broadband (eMBB), massive Machine-Type Communications (mMTC) and Ultra-Reliable Low Latency Communications (URLLC)\cite{8101523}. URLLC, which is a key design aim in 5G, supports short packet transmissions with ultra-high reliability and low latency in order of milliseconds which is compatible with different applications such as road safety, industrial control, and vehicle to vehicle communications, where a failure leads to drastic subsequences~\cite{7529226},~\cite{popovski2017ultra}.

During the past couple of years, short packet transmission has gained much attention from academia and industry since the majority of theoretical works assume infinite blocklenghts (IFB). %For instance, an overview of some coding schemes under FB coding which may be utilized in 5G is provided in~\cite{iscan2016comparison}. They indicate new coding schemes where the decoders are computationally sophisticated, reduce the performance loss of communications under FB regime. Supporting URLLC requirements improve the security, functionality and the ability of the interaction between different types of communication units such as human-to-human, human-to-machine or machine-to-machine which creates novel business models and applications~\cite{7786123}. Critical requirements of $5$G such as data rate, energy, and cost issues are discussed in more details in~\cite{popovski2017ultra,andrews2014will,7736615}.  
Stringent latency and reliability requirements of URLLC with FB may be hard to achieve with traditional communication techniques. Our prior work in~\cite{parisa2017} and~\cite{nouri2017ultra} shows that cooperative relaying is a potential solution to guarantee ultra-reliability and meet URLLC requirements. %Relaying is a well-known technique which improves the performance of communication links by decreasing the multi-path fading through reducing the path loss and exploiting the spatial diversity\cite{8329625}. In cooperative relaying, auxiliary nodes transfer the message from the source to destination. In such networks, destination receives independent replicas of a message without installing collocated antennas at the source/destination which reduces the transmit power while increasing the capacity, providing better signal quality and achieving diversity against fading~\cite{mansourkiaie2015cooperative}. %Decode-and-forward (DF) protocol is the most common relaying scheme where the relay decodes, encodes and retransmit the message to the destination~\cite{zimmermann2005performance}.
A large body of work has studied, the impact of cooperative diversity in several systems and channel models. For instance, a comprehensive study of existing cooperative schemes is provided in~\cite{mansourkiaie2015cooperative}. %Authors in~\cite{swamy2015cooperative}, address the reliability and latency requirements by taking the advantage of cooperative diversity.  
However, the impact of cooperative relaying on the performance of communications is generally studied under the ideal assumption of Shannon channel coding theorem where there is no bound on the communication blocklength, in other words, the system works under IFB regime and error-free communication is theoretically possible \cite{8329625},~\cite{hu2015performance}. However, in~\cite{polyanskiy2010channel} the authors show the existence of a significant performance loss with finite blocklength (FB) compared to the Shannon limit. On the other hand, most wireless communication systems apply periodic pilot signals to estimate the channel before data detection.  Channel estimation overhead is one of the key characteristics of wireless systems which greatly affect the minimum transmission delay. Channel estimation quality enhances with the number of pilot symbols; however, the increase in the length of training reduces the effective data rate and wastes resources that could be used for data symbols~\cite{lozano2010optimum}.
%Taking this into account in addition to the stringent latency and reliability requirements with short packet transmissions, we examine the impact of pilot overhead on the performance of URLLC in cooperative and non-cooperative transmissions since the sizes of message payload and pilot overhead are often of the same order of magnitude and unsuccessful design of the pilots increases the performance loss.
%\subsection{On the Impact of Pilot Overhead - Related works } 
Different factors such as mobility, environment and multipath fading, which change the channel conditions during the transmission time, lead to random fluctuations in the power of the received signal and complicate the channel estimation task~\cite{gursoy2009capacity}. %With imperfect channel state information (CSI), it is common to assume that the channel estimation error is a Gaussian random variable with zero mean. Hence, errors in the channel estimation restrict the channel capacity in addition to the channel noise~\cite{haghighi2010energy}.

The performance of wireless communications under imperfect CSI has been extensively studied in the literature. For example,~\cite{bao2013effect} study the performance of cognitive multihop relay networks. Authors propose a backoff control power technique to deal with such imperfection due to the fact that the efficiency of such networks is bounded by the channel uncertainty. The impact of the training length on the performance of high mobility systems is studied in~\cite{sun2014maximizing}. Authors quantify the optimal number of pilots in such a way that the lower bound of the spectral efficiency is maximized.
%In the context of FB, authors in~\cite{DBLP:journals/corr/HuGS16}, examine the impact of imperfect CSI on the performance of a relaying network. They propose dynamic and constant SNR weight schemes in order to enable the source to determine the appropriate coding rate per transmission period. 
Authors in~\cite{schiessl2018delay} provide the error probability in closed form, considering the joint impact of imperfect CSI at the transmitter and FB coding, where the performance of wireless systems under fixed rate and rate adaptation schemes is investigated. Systems with rate adaptation outperform systems working at a fixed rate, although the former require more channel uses for the training sequence. However, the work in~\cite{schiessl2018delay} does not apply to URLLC as the rate adaptation technique causes random delay due to the variation of data rate from the slot to slot and data stays in a queue for some random time. Authors in~\cite{mousaei2017optimizing}, study the pilot overhead optimization in an ultra-reliable short packet point-to-point communication system in terms of achievable rates. They compare the optimal overhead in block fading and continuous fading channels subject to FB coding and IFB coding. They indicate that the optimal pilot overhead is almost similar in both fading channels since block fading channels are a specific case of continuous fading models and they can be unified in the context of training based communications~\cite{lozano2010optimum}. Moreover, the optimal size of pilot estimation decreases with transmit power and increases with Doppler frequency. 

Clearly, there is a need to study the impact of pilot overhead in the context of URLLC. The performance and design of wireless systems highly depend on the channel conditions which motivates us to examine the impact of channel knowledge in URLLC. Differently from our previous works \cite{parisa2017, nouri2017ultra} which assume ideal channel estimation, herein we aim to investigate the impact of channel estimation overhead on the performance of decode-and-forward (DF) cooperative communications in URLLC.  
The contributions of this work are: $i)$ we extend our previous works in~\cite{parisa2017} and~\cite{nouri2017ultra} by studying the impact of channel estimation on cooperative communications over Rayleigh fading channels subject to two distinct power constraints; peak power constraint (PPC) and average power constraint (APC) $ii)$ we indicate the impact of PPC factor on URLLC performance metrics such as minimum latency and goodput in cooperative and non-cooperative transmissions with optimal training overhead, $iii)$ we illustrate the impact of peak power constraint factor on the training sequence and derive the optimal number of pilots in closed form.

The rest of this paper is organized as follows. Section \ref{sc:system_model} presents the system and channel models. Section \ref{cooperative} discusses the outage probability of cooperative communications with FB, and Section \ref{numerical results} presents the simulation results. Section \ref{conclusion} concludes the paper.
\vspace{-2mm}
\section{System and Channel Models} \label{sc:system_model}
\begin{figure}[t!]
	\centering
	\includegraphics[width=\columnwidth]{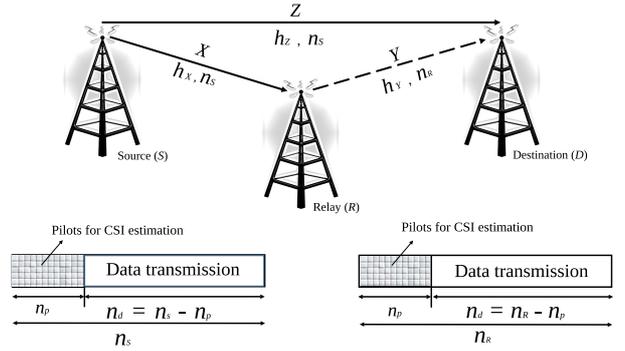}
	\vspace{-3mm}
	\caption{System model and packet structure of in-band pilot overhead for relaying scenario. Broadcasting and relaying phases are indicated with solid and dashed lines, respectively.}
	\label{fig:System Model}
\end{figure}

Consider a DF relaying scenario that consists of a source $S$, a destination $D$ and a relay $R$. The direct link $S$-$D$, backhaul link $S$-$R$ and relay link $R$-$D$ are respectively denoted by $Z$, $X$ and $Y$, and each transmission takes $n_S$ channel uses in the broadcasting phase and $n_R$ channel uses in the relaying phase where $n_p$ channel uses out of the total channel uses is allocated to channel estimation as illustrated in Fig.~\ref{fig:System Model}. The distance of the direct link is normalized as $d_{SD}\!=\!1$ and $R$ can move between $S$ and $D$; hence, $d_{SR}\!=\!\beta$ and $d_{RD}\!=\!1-\beta$. In this scenario, in the broadcasting phase $S$ sends the message to $D$ and $R$; in the relaying phase, if it decoded the message correctly, $R$ forwards the message to $D$. The received signal in the broadcasting phase at $D$ and $R$ are $y_1\!=\!h_Zx\!+\!w_Z$ and $y_2\!=\!h_Xx\!+\!w_X$, respectively, and if $R$ collaborates with $S$, the received signal at $D$ is $y_3\!=\!h_Yx\!+\!w_Y$, where $x$ is the transmitted signal with power $P$ and $w_i$, $i\!\in\!\{X,Y,Z\}$ is noise, modeled as Gaussian where $w_i\!\sim\!{\cal C} {\cal N} (0,1) $. The Rayleigh fading channels in $S$-$D$, $S$-$R$ and $R$-$D$ links are denoted by $h_Z$, $h_X$ and $h_Y$, respectively where $h_i$ is a zero mean circularly symmetric complex Gaussian random variable with unit variance $\sigma^2\!=\!1$. In a DF-based cooperative transmission, the instantaneous SNR for each link depends on the total power constraint $P\!=\!P_S\!+\!P_R\!=\!\eta P\!+\!(1-\eta)P$, and are given by $\Omega_Z\!=\!\eta d_{SD}^{-\alpha} P|h_Z|^2$, $\Omega_X\!=\!\eta d_{SR}^{-\alpha} P|h_X|^2$ and $\Omega_Y\!=\!(1\!-\!\eta) d_{RD}^{-\alpha} P|h_Y|^2$, where $\eta$ is the power allocation factor between S and R and $\alpha$ is the path loss; thus, the average SNR in each link are $\gamma_Z\!=\!\eta d_{SD}^{-\alpha} P$, $\gamma_X\!=\!\eta d_{SR}^{-\alpha} P$ and $\gamma_Y\!=\!(1-\eta) d_{SD}^{-\alpha}P$.

%\subsection{In-Band Pilot Estimation Over Fading Channels}
According to Fig. \ref{fig:System Model}, in-band pilot transmission consists of a training phase and a data transmission phase. The received signal in the training phase is $y_{t_i}\!=\!h_ix_{t}\!+\!w_{i}$ where $i \!\in\!\{X,Y,Z\}$ and $||x_t||^2\!=\!n_pP$, where $n_p$ is the number of pilot symbols. We consider the use of the minimum mean square error (MMSE) approach to estimate the channel coefficient as $\hat{h_i}\!=\!\operatorname{E}\{h_i|y_{t_i}\}=\frac{\sigma^2}{\sigma^2 ||x_t||^2+1}x^{\dag}_t y_{t_i}$~\cite{gursoy2009capacity}, where
\begin{equation}\label{hat}
\hat{h_i}\sim {\cal CN} \left(0,\dfrac{\sigma^4||x_t||^2}{\sigma^2||x_t||^2+1}\right) = {\cal CN} (0,\sigma_{\hat{h_i}}^2),
\end{equation} 
and the error in the channel estimation is assumed to be Gaussian as
%\vspace{-2mm}
\begin{equation}\label{tilde}
\tilde{h_i}\sim{\cal CN}\left(0,\dfrac{\sigma^2}{\sigma^2||x_t||^2+1}\right)={\cal CN}(0,\sigma_{\tilde{h_i}}^2).
\end{equation}

With the knowledge of the channel condition at the receiver, the received signal can be written as $y_{d_i}\!=\!\hat{h_i}x_{d}\!+\!\tilde{h_i}x_{d}\!+\!w_{i}\!=\!\hat{h_i}x_{d}\!+\!w_{{\text{eff}}_i}$, where additive noise and residual channel estimation error are combined in $w_{{\text{eff}}_i}$~\cite{gursoy2009capacity}. The main difference between the received signals in the training and data transmission phases is that the channel in the latter phase is known to the receiver. According to the orthogonality property of MMSE estimations, where $\sigma_{\hat{h_i}}^2\!=\!1\!-\!\sigma_{\tilde{h_i}}^2$, the effective SNR $\gamma_{\text{eff}}$ seeing at the receiver is~\cite{mousaei2017optimizing}
\begin{equation}\label{effectiveSNR}
\gamma_{\text{eff}}\!=\! \dfrac{\sigma_{\hat{h_i}}^2\sigma_d^2}{1\!+\!\sigma_{\tilde{h_i}}^2\sigma_d^2}\!=\! \dfrac{\sigma_d^2 (1\!-\!\sigma_{\tilde{h_i}}^2)}{1\!+\! \sigma_{\tilde{h_i}}^2\sigma_d^2}\!=\!\dfrac{1\!+\!\sigma_d^2}{1\!+\!\sigma_{\tilde{h_i}}^2\sigma_d^2}\!-\!1,
\end{equation}
where $\sigma_d^2$ is the data mean power~\cite{mousaei2017optimizing}. Thus, we should find the optimal number of pilot symbols which maximizes $\gamma_{\text{eff}}$ or, in other words, minimizes the mean square error $\sigma_{\tilde{h_i}}^2$. 

In the following, we examine the performance of such a communication under two distinct policies: APC and PPC. %First, we start with APC on the pilot and data symbols, where a single pilot with high power is sent in the training phase, while PPC allocates an optimal fraction of resources to the training sequence to meet URLLC requirements.
\vspace{-2mm}
\subsubsection{Average Power Constraint (APC)}\label{APC}
We consider an average power constraint as $\operatorname{E}\{|x|^2\}\leq nP$, where the power budget is distributed among data and pilot symbols. %We can maximize the effective SNR by optimizing training parameters such as the training length and pilot power. In terms of channel estimation performance, using $n_p$ pilot symbols is the same as sending a single pilot with power scaled by $n_p$. 
This scenario could be optimal if we send a single pilot with high power since increasing the size of the pilot sequence decreases the outage probability at the cost of an increased delay. %However, excessively high pilot power may be prohibitive in practical systems.
We allocate an optimal fraction of the power budget $\psi$ to the pilot symbol, where %$|x_t|^2\!=\!\psi nP$ and $|x_d|^2\!=\!\frac{(1-\psi)nP}{n-1}\text{I}$, where $\psi\!=\!\sqrt{\xi(\xi\!+\!1)}\!-\!\xi$ and $\xi\!=\!\frac{nP+(n-1)}{Pn(n-2)}$~\cite{gursoy2009capacity}. Hence,
the effective SNR is calculated according to ~\cite[\S 15]{gursoy2009capacity} as
\begin{equation}\label{SNReffective}
\gamma_{\text{eff}_\text{APC}}\!=\! \dfrac{nP\left(1\!+\!\operatorname{D}\!-\!\sqrt{\operatorname{F}}\right)\left(\sqrt{\operatorname{F}}\!-\!\operatorname{D}\right)}{(n\!-\!2)\sqrt{\operatorname{F}}}
,\!
\end{equation}
where $\operatorname{D}\!=\!n\!+\!nP\!-\!1/(n\!-\!2)nP$ and $\operatorname{F}=\frac{(n\!-\!1)(n^2P(1\!+\!P)\!+\!n\!-\!1)}{(n\!-\!2)^2n^2P^2}$. Hence, increasing the power of pilot or increasing the number of pilots decreases the power to be used in data.
\vspace{2mm}
\subsubsection{Peak Power Constraint (PPC)}
Since pilot power subject to APC increases with blocklength and in practice, increasing the pilot power beyond a limit is in general not possible, we consider a peak power constraint on the pilot symbols as $|x_t|^2\!\leq\!\kappa P$ in addition to the average constraint, where $\kappa$ is the peak power constraint factor, and PPC activates if $\psi nP>\kappa P$~\cite{gursoy2009capacity}. The optimal number of pilot symbols is the one that maximizes the effective SNR. Thus, considering $||x_t||^2\!=\!\kappa n_pP$, the effective SNR subject to PPC is
\begin{align}\label{SNReffective_PPC}
\gamma_{\text{eff}_\text{PPC}} = \dfrac{n_p \kappa (n-n_p\kappa)P^2}{(n-n_p\kappa+(n-n_p)n_p\kappa)P+n-n_p},
\end{align}
where the optimum number of pilot symbols with unbounded $\kappa$ would be $1$, but the optimum transmit power would be too large~\cite{gursoy2009capacity}.
The optimal number of pilot symbols which maximizes $\gamma_{\text{eff}_\text{PPC}}$ obtains through the numerical derivative of (\ref{SNReffective_PPC}) with respect to $n_p$, as
\begin{align}\label{Np}
\begin{split}
\dfrac{\partial \gamma_{\text{eff}_\text{PPC}}}{\partial n_p} &= (\kappa^2P^2\!+\!\kappa^3P^3\!-\!n\kappa^3P^3+n\kappa^2P^3)n_p^2\!-\!(2n\kappa^2P^2\\
&+\!2n\kappa^2P^3)n_p + (n^2\kappa P^2\!+\!n^2\kappa P^3) = 0,
\end{split}
\end{align}
where the positive root of the polynomial function is the optimal size of the training sequence. Hence,
%\begin{figure*}[!t]
%	\centering
%	\vspace{-5mm}
%	\begin{equation}\label{Np}
%	\partial \gamma_{\text{eff}_\text{PPC}}/\partial n_p = (\kappa^2P^2\!+\!\kappa^3P^3\!-\!n\kappa^3P^3+n\kappa^2P^3)n_p^2\!+\!(\!-\!2n\kappa^2P^2\!-\!2n\kappa^2P^3)n_p + (n^2\kappa P^2\!+\!n^2\kappa P^3) = 0.
%	\end{equation}
%	\hrule
%\end{figure*}
\begin{equation}
n_{p_\text{opt}}\!=\!\frac{-\!\operatorname{B}\!\pm\!\sqrt{\operatorname{B}^2-4\operatorname{A}\operatorname{C}}}{2\operatorname{A}},
\end{equation}
where $\operatorname{A}=\kappa^2P^2\!+\!\kappa^3P^3\!-\!n\kappa^3P^3+n\kappa^2P^3$, $\operatorname{B} = \!-\!2n\kappa^2P^2\!-\!2n\kappa^2P^3$ and $\operatorname{C}= n^2\kappa P^2\!+\!n^2\kappa P^3$.

In this work, we assume that the source immediately transmits with a fixed coding rate and rely on the average SNR	of each link due to the timing constraint where we do not 	have a large tolerance for latency with respect to the whole transmission time. Optimizing the number of pilots according to the performance of the relay improves the performance of the communication at the cost of higher latency and more sophisticated implementation procedure.
\vspace{-2mm}
\section{Performance analysis of Relaying under Finite Blocklength}\label{cooperative}
%\subsection{Single-Hop Communication with Finite Blocklength}
In single-hop communication, first $k$ information bits are encoded into a codeword of $n$ symbols, which is forwarded to a decoder via the wireless channels. Afterwards, the decoder maps the channel outputs into an estimate of the information bits. Hence, the coding rate is given by the ratio of the information bits $k$ to the total number of channel uses $n$, $\cal R$$\!$$=\!k/n$~\cite{7529226}. Authors in \cite{polyanskiy2010channel}, indicate that assuming the coding rate as $k/n$ under the FB regime does not guarantee the outage probability as small as desired, due to the ultra-reliable constraint, $n$ should be very large and so ${\cal{R}}\rightarrow 0$. Hence, the maximum coding rate $\cal R^*$$(n,\epsilon)$ in bits per channel uses (bpcu) for single-hop communication with finite blcoklength $n$, outage probability $\epsilon$ and the average SNR $\gamma$ is
\vspace{-2mm}
\begin{equation} \label{eq:maximum rate}
{\cal R}^*(n,\epsilon)\!=\!C(\gamma)\!-\!\sqrt{\dfrac {V(\gamma)}{n}}\operatorname{Q}^{-1}(\epsilon)\log_{2}\operatorname{e},
\end{equation}
where, $C(\gamma)\!=\!\log_{2}(1\!+\!\gamma)$ and the channel dispersion is $V(\gamma)\!=\!\gamma (2\!+\!\gamma) \big/(1\!+\!\gamma)^2$, where $\frac{1}{n}\!\sum_{i}^{n}|x_{i}|^2\!\leq\!\gamma$ holds in AWGN channels~\cite{7529226}.  According to (\ref{eq:maximum rate}), the maximum achievable coding rate with FB coding increases by blocklength $n$, however, there would be a performance gap compared to the Shannon capacity as shown in \cite[Fig. $1$]{8329625}.
According to (\ref{eq:maximum rate}), outage probability of any block fading channel is given by~\cite{7529226}

\begin{equation} \label{outage_fading}
\epsilon\approx \operatorname{E}\Bigg[\operatorname{Q}\Bigg(\sqrt{n}\frac{C(\gamma|h|^2)-{\cal R}^{*}(n,\epsilon) }{\sqrt{V(\gamma|h|^2)}}\Bigg)\Bigg],\\
\end{equation}
where the expectation is taken over the channel distribution. By keeping in mind the importance of using FB coding to overcome the challenges that machine-type communications will face in $5$G networks, authors in \cite{mary2016finite} also mention the necessity of using FB coding with block fading channels since IFB coding overestimates resource allocation under block fading channels.

%\subsection{Outage Probability in Closed-form}
The outage probability in (\ref{outage_fading}) does not have a closed-form expression, but it can be tightly approximated as%[\S 13 and 14]
\begin{align}\label{outage_fading_linear}
\!\!\!\!\epsilon\!=\!1\!-\!\dfrac{\zeta}{\sqrt{2\pi}}\exp(-\theta)\!\bigg[\!\exp\!\bigg(\!\sqrt{\dfrac{\pi}{2\zeta^2}}\!\bigg)\!\!-\!\exp\!\bigg(\!\!-\!\sqrt{\dfrac{\pi}{2\zeta^2}}\!\bigg)\!\!\bigg]\!,
\end{align}
where $\theta\!=\!\frac{2^{\cal R}-1}{\gamma}$, $\zeta\!=\!\gamma\mu\sqrt{2\pi}$ and $\mu\!=\!\sqrt{\dfrac{n/2\pi}{\operatorname{e}^{2\cal R}-1}}$~\cite{6888474}. The accuracy of the approximated outage probability in (\ref{outage_fading_linear}) is illustrated in \cite[Fig. $3$]{Nouri2018}. It should be noted that the effect of channel estimation errors is seen in the effective SNR $\gamma_{\text{eff}}$ which is then used in the outage expression.
\vspace{2mm}
\subsection{Decode-and-Forward (DF)}\label{sec:MRC}
Transmissions from $S$ and $R$ are coherently combined at the receiver. The instantaneous SNR after $S$ and $R$ transmissions is $\gamma_{W}\!= \!\gamma_{Z}\!+\!\gamma_{Y}$~\cite{alves2012throughput}. Outage probability is given by~\cite{parisa2017}
\vspace{-2mm}
\begin{equation}\label{eq:MRC_outage}
\epsilon_\text{DF} =\epsilon_{Z}\left(\epsilon_{X}\!+\!\left(1\!-\!\epsilon_{X}\right)\frac{\epsilon_\text{SRD}}{\epsilon_{Z}}\right),
\end{equation}
where $\epsilon_\text{SRD}$~\cite[\S 6]{parisa2017} is the outage probability after maximum ratio combining of the transmissions from $S$ to $D$. The ratio $\frac{\epsilon_\text{SRD}}{\epsilon_{Z}}$ comes from the conditioning of $\epsilon_\text{SRD}$ on the fact that the transmission from $S$ to $D$ failed. Notice that $\epsilon_{Z}$ and $\epsilon_{X}$ are calculated according to (\ref{outage_fading_linear}), where $\zeta$ is updated with
$P_S = \eta \gamma_{\text{eff}}$, $P_R = (1-\eta)\gamma_{\text{eff}}$ and $\mu$ with $n=n_S$, $n=n_R$.
\subsection{Direct Transmission}
In point-to-point communications, source sends the message directly to the destination. The outage probability is calculated as in (\ref{outage_fading_linear}), where $\gamma$ and $\mu$ are updated with $P\!=\!P_S$ and $n\!=\!n_S$, respectively.
\vspace{1mm}
\section{Numerical Results}\label{numerical results}
In the numerical results, unless stated otherwise, we assume ${\cal {R}} = 0.5$ (bpcu), $\kappa\!=\!3$, maximum transmit power per link as $10$ dB and that $R$ is between $S$ and $D$, with $\beta\!=\!0.5$, $\eta\!=\!0.5$, and path loss as $\alpha\!=\!4$. The assumed ultra-reliable region (URR), where the outage probability is $10^{-3}$ and $99.9 \%$ reliability
is guaranteed, is shaded in gray in the following plots, and its most loose constraint with $99.9 \%$ reliability is denoted with a red line.
%\subsection{On the Impact of In-Band Pilot Estimation}

\begin{figure}[!t]
	\centering
	\includegraphics[width=\columnwidth]{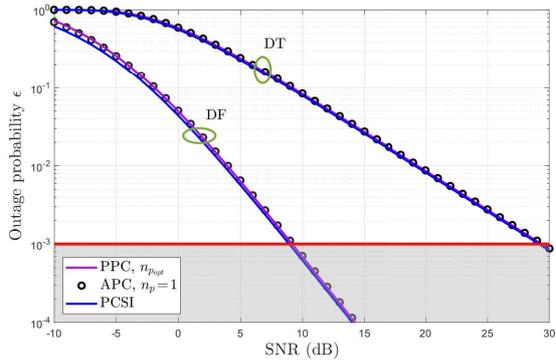}
	\vspace{-3mm}
	\caption{Comparing the performance of cooperative and non-cooperative transmissions under APC and PPC.}\label{fig:outage}
\end{figure}

Fig. \ref{fig:outage} compares the performance of training-based cooperative and non-cooperative schemes subject to APC and PPC, with the case of perfect CSI (PCSI). Under APC, we send a single pilot with high power where the pilot power increases by $n$, while under PPC, we send an optimal number of pilots based on the maximum transmit power and value of $\kappa$. Performance of both cooperative and non-cooperative transmissions subject to APC and PPC is highly close in the entire range. In communications subjected to APC, we have to allocate about half of the power budget to a single pilot at low SNR regime, while with PPC, we spread that amount of power among several pilots, and so, we improve the channel estimation while the pilot power does not go beyond the power threshold. %Therefore, in the following figures, we study PPC policy in more details.
In addition, we can clearly see the superiority of cooperative transmissions over non-cooperative one. The cooperative technique provides higher reliability with lower transmit power compared to the point-to-point transmission.

\begin{figure}[!t]
	\centering
	\includegraphics[width=\columnwidth]{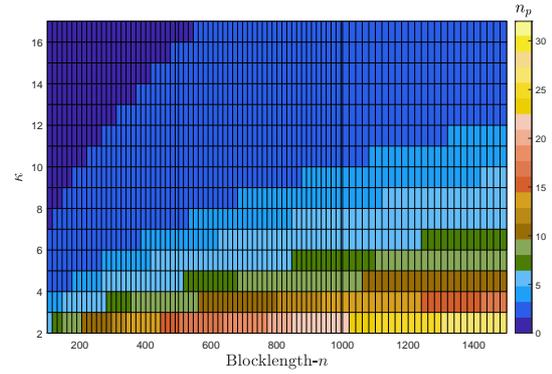}
	\vspace{-3mm}
	\caption{Impact of peak power constraint factor ($\kappa$) on the length of training sequence ($n_p$) in terms of channel uses.}\label{fig:kappaimpact}
\end{figure}

In Fig. \ref{fig:kappaimpact}, we indicate the impact of $\kappa$ on the optimal length of the training sequence. It can be clearly seen that the number of pilot symbols decreases by increasing $\kappa$ to keep the pilot power under the power threshold, and so, if the blocklength increases for a constant value of $\kappa$, we have to allocate more channel uses to the training sequence. On the other hand, $\gamma_{\text{eff}}$ highly depends on $n_p$ and $\kappa$ and decreases by large values of $\kappa$ with short packet transmissions. Hence, channel estimation and  $\gamma_{\text{eff}}$ are extremely affected by $\kappa$ .
% 
%\subsection{URLLC Performance Metrics}
%\subsubsection{Minimum Required Latency}

\begin{figure}[!t]
	\centering
	\includegraphics[width=\columnwidth]{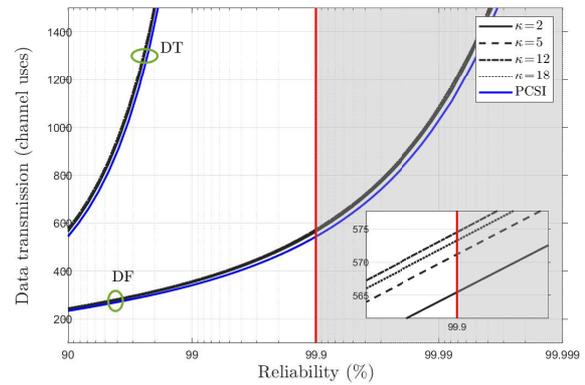}
	\vspace{-3mm}
	\caption{Minimum required channel uses for data transmission phase with maximum transmit power as $20$ dB.}\label{fig:latency}
\end{figure}

\begin{figure}[!t]
	\centering
	\includegraphics[width=\columnwidth]{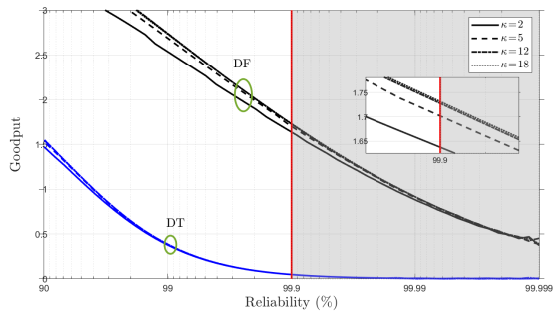}
	\vspace{-3mm}
	\caption{Goodput as a function of reliability with maximum transmit power as $20$ dB.}\label{fig:goodput}
\end{figure}

In the next following figures, we examine two URLLC performance metrics, namely, minimum latency and goodput as a function of reliability. In both Fig. \ref{fig:latency} and Fig. \ref{fig:goodput}, the choice of the optimal length of data transmission phase and training sequence is made in such way that minimizes the outage probability in a specific interval of interest as $10^{-5}\!\leq\!\epsilon\!\leq\!10^{-1}$, giving the minimum latency and maximum goodput. We resort to Matlab function "$fmincon$" and use interior point algorithm as detailed in \cite{Waltz2006} to numerically solve the nonlinear optimization problem as we do not focus on the proposal of a particular solution. The optimal size of the training sequence is obtained through the numerical derivative of (\ref{SNReffective_PPC}) with respect to $n_p$. 

Fig. \ref{fig:latency} compares the reliability and minimum channel uses needed for data transmission phase with optimal training length and different values of $\kappa$. Delay $\delta$ is equal to the symbol time $T_s$ multiplied by the blocklength of data transmission. For example, future releases of LTE foresee a minimum symbol period of $T_s\!=\!8.33$$\mu$s~\cite{yilmazultra}. In this current work, we do not consider the impact of processing delay and channel coding at the relay  in our analysis. However, as a future work we will account for processing delay and practical channel coding schemes for URLLC.
In cooperative transmission with maximum transmit power as $20$ dB per link and $\delta\!=\!4.7$ ms, $99.9 \%$ reliability is feasible with $n_p\!=\!11$ (channel uses) and $\kappa\!=\!2$, while reliability slightly decreases with larger $\kappa$. Reliability improves to $99.99 \%$ with $\kappa\!=\!2$, $n_p\!=\!15$ (channel uses) at the cost of $\delta\!=\!9$ ms. Moreover, by decreasing the maximum transmit power the latency increases. For example, with maximum transmit power per link as $10$ dB, latency increases to $6.29$ ms and $17.2$ ms with reliability as $99.9 \%$ and $99.99 \%$, respectively. Hence, according to our system design requirements, we can find the trade-off between the reliability and latency. Moreover, the results show that non-cooperative scheme is not able to cope with the stringent reliability requirement and needs a large tolerance of latency to work under URR, and so, cooperative communications are needed for the optimal performance under URR. 
%
%\subsubsection{Goodput with Rate Adaptation}

Another performance metric of URLLC is the goodput which indicates how much data is successfully received at the destination. The total amount of data is transmitted over $n$ channel uses where the transmission might fail with probability of $\epsilon$~\cite{schiessl2018delay}. The normalized goodput with respect to the total number of channel uses allocated to packet transmission is $GP = \left(1-\frac{n_p}{n}\right){\cal {R}} \left(1-\epsilon\right).$
 
%We maximize the goodput as follows

%\begin{equation*}
%\begin{aligned}
%\centering
%& \underset{}{\text{Maximize}}
%& &  GP \left(\epsilon,n,n_p\right), \\
%& \text{subject to}
%& &100\leq n \leq 10000\\
%& & & n_p > 1,\\
%& & & \epsilon \leq \epsilon_{\text{th}}.
%\end{aligned}
%\end{equation*} 

%This problem is equivalent to minimizing the outage probability with respect to the blocklength $n$ and training length $n_p$ in an interval of interest as $10^{-5} \leq \epsilon \leq 10^{-2}$. We resort to Matlab function $f_{mincon}$ and use interior point algorithm as detailed in \cite{Waltz2006} to numerically solve the nonlinear optimization problem as we do not focus on the proposal of a particular solution. The optimal size of training sequence is obtained through the numerical derivative of (\ref{SNReffective_PPC}) with respect to $n_p$. 

From Fig. \ref{fig:goodput}, we observe that the goodput decreases by increasing the reliability. In this optimization problem, we find the optimal blocklength $n$ and training length $n_p$ in such a way to guarantee a certain reliability. Hence, the coding rate adopts for each $\epsilon_{\text{th}}$ in the optimization problem and consequently the goodput decreases as the coding rate monotonically decreases, however, with fixed coding rate, the goodput increases but we can not guarantee the intended reliability. In addition, we can note that by allocating more power to the pilot symbols (larger $\kappa$),  goodput increases due to the fact that the total blocklength $n$ and the number of channel uses allocated to the data transmission increases while the training length decreases and $\xi = n_p/n$ reduces faster compared to the smaller values of $\kappa$. It can be noted that the goodput analysis also makes it possible to conclude the throughput's behavior in URLLC. The throughput is the average rate seen by the destination. The throughput as well the goodput decreases at high reliabilities since we can not concurrently guarantee both high throughput and ultra-high reliability for URLLC, i.e in order to have URLLC we sacrifice the goodput and throughput. Clearly, ultra-high reliability is feasible at the expense of throughput loss and vice versa. All in all, according Fig. $4$ and Fig. $5$, we can find out the expected goodput for specific latency and reliability requirements.
\vspace{2mm}
\section{Conclusions and Final Remarks}\label{conclusion}
We investigated the impact of in-band pilot overhead on the performance of cooperative and non-cooperative communications subject to APC and PPC policies. Numerical results show that the performance of in-band pilot transmissions under PPC is pretty close to APC. %Moreover, we indicated the impact of power allocation factor and blocklength on the optimal training length. 
The training length decreases by allocating more power to the pilot symbols and increases by increasing the blocklength. Furthermore, by allocating more power to the pilot symbols (larger $\kappa$) for a specific latency requirement, reliability decreases as the training length decreases, and so the channel estimation deteriorates. On the other hand, increasing the pilot power improves the required goodput to guarantee a certain reliability. In addition, we showed the importance of cooperative communications over DT, where diversity technique could be a feasible solution to meet URLLC requirements. We should keep in mind that short message communications must confine the training sequence length; thus, PPC on pilot symbols, provides an optimal transmission power to support maximum achievable data transfer over the whole blocklength, which is more appealing in the context of URLLC. 
  
\vspace{2mm}
\section*{Acknowledgments}
This work has been partially supported by the Academy of Finland 6Genesis Flagship (grant no. 318927), Finnish Funding Agency for Technology and Innovation (Tekes), Huawei Technologies, Nokia and Anite Telecoms, Academy of Finland (under Grant no. 307492), and CNPq (Brazil).

%\begin{figure}[!t]
%	\centering
%	\includegraphics[width=\columnwidth]{6.pdf}
%	\vspace{-3mm}
%	\caption{ Impact of relay positioning on the performance cooperative transmission subject to PPC.}\label{fig:relaypositioning}
%\end{figure}
%\section*{Acknowledgment}\unskip~\cite{}


%\bibliographystyle{IEEEtran}
\vspace{-2mm}
\bibliographystyle{IEEEtran}
%\bibliography{ref}

\end{document}